\documentclass[preprint,12pt]{elsarticle}

\newcommand{\norm}[1]{\left\lVert#1\right\rVert}
\newcommand{\bm}[1]{\boldsymbol{#1}}
\newcommand{\inv}[1]{{#1}^{-1}}
\usepackage{mathtools}
\usepackage{hyperref}

\usepackage{amssymb}
\usepackage{amsmath}

\usepackage{array}
\usepackage{multirow}
\newcolumntype{C}{>{\centering\arraybackslash}m}
\newcolumntype{L}{>{\arraybackslash}m}
\usepackage{seqsplit}
\renewcommand{\seqinsert}{\ifmmode\allowbreak\else\-\fi} % adds hyphen

\newcommand{\one}[1]{{#1}} % to highlight changes for reviewer one
\newcommand{\two}[1]{{#1}} % to highlight changes for reviewer two
\newcommand{\three}[1]{{#1}} % to highlight our own improvements

\begin{document}

% \listoftodos

\begin{frontmatter}

\title{Domain decomposition of the modified Born series approach for large-scale wave propagation simulations}
\author[label1]{Swapnil Mache\fnref{label2}}
\author[label1]{Ivo M. Vellekoop\corref{cor1}}
\cortext[cor1]{Corresponding author. \textit{E-mail address:} i.m.vellekoop@utwente.nl (I. M. Vellekoop)}
\affiliation[label1]{
    organization={Biomedical Photonic Imaging Group, Faculty of Science and Technology, University of Twente}, 
    addressline={P.O. Box 217}, 
    city={Enschede}, 
    postcode={7500 AE}, 
    % state={},
    country={The Netherlands}}%
\fntext[label2]{Previous address: Rayfos Ltd., Winton House, Winton Square, Basingstoke, United Kingdom}
\begin{abstract} % max 250 words
The modified Born series (MBS) is a fast and accurate method for simulating wave propagation in complex structures. \two{In the current implementation of the MBS, the simulation size is limited by the working memory of a single computer or graphics processing unit (GPU). Here, we present a domain decomposition method that enhances the scalability of the MBS by distributing the computations over multiple GPUs, while maintaining its accuracy, memory efficiency, and guaranteed monotonic convergence. With this new method, the computations can be performed in parallel, and a larger simulation size is possible as it is no longer limited to the memory size of a single computer or GPU. We show how to decompose large problems over subdomains and demonstrate our approach by solving the Helmholtz problem for a complex structure of $3.28\cdot 10^7$ cubic wavelengths ($320 \times 320 \times 320$ wavelengths) in just $45$ minutes with a dual-GPU simulation.}
\end{abstract}
%
%% Keywords
\end{frontmatter}
%
%
%%%%%%%%%%%%%%%%%%%%%%%%%% body %%%%%%%%%%%%%%%%%%%%%%%%%%
%
\section{Introduction}
\label{sec:introduction}
\noindent Wave propagation simulations have many applications, ranging from nano\-photonics to geophysics. Unfortunately, computing accurate solutions to wave equations in large heterogeneous media is highly time-consuming. A great number of numerical methods are available for solving wave equations in heterogeneous media \cite{jin2015theory}. \one{The modified Born Series (MBS) \cite{osnabrugge2016convergent, osnabrugge2021ultra} is one of the fastest and most accurate methods \cite{kruger2017solution, lee2022inverse}}.

\three{The MBS is a frequency domain method, where the structure is represented on a regular grid. This grid-based approach allows arbitrarily complex structures to be defined and enables an efficient computation of the iteration steps. Transient wave phenomena, which are especially important in sound and seismic waves, can be simulated by solving for each frequency component individually, as is common in frequency-domain methods \cite{rappaport1991fdfd, lui1999direct, bayliss1985numerical, thompson2006review, plessix2007helmholtz, wang20113d}. This approach has the advantage of being able to accurately describe physical dispersion effects (i.e., wavelength-dependent propagation speed).}

\three{A key benefit of the MBS is that it does not rely on finite difference approximations. This sets it apart from widely used grid-based methods such as the finite-difference time-domain (FDTD) \cite{yee1966numerical, taflove1995computational, oskooi2010meep, nabavi2007new} and pseudospectral time-domain (PSTD) \cite{liu1997pstd, fornberg1987pseudospectral, wise2021pseudospectral} that rely on finite difference approximations for the spatial (FDTD) and temporal (FDTD and PSTD) derivatives. These approximations introduce errors that accumulate as the wave propagates \cite{kruger2017solution, liu1997pstd, zheng2001numerical, jin2010concluding, dwarka2021pollution}. Although it is possible to reduce these errors by decreasing the voxel size (for FDTD) and the time step (for FDTD and PSTD) \cite{osnabrugge2016convergent, johnson2013advances, gibson2021method, he2023euv, he2024modified}, this comes at a correspondingly steep cost in memory usage and computation time while giving a limited benefit; for example, even at a sampling of 10 points per wavelength, the phase velocity error of FDTD is a few percent \cite{heh2009dispersion}, resulting in a phase error of over $\pi$ after propagating roughly a dozen of wavelengths. Similarly, accumulating errors in the time derivative limit PSTD to simulations far less than 100 wavelengths, even at a time step of 1/10 per propagated wavelength \cite{osnabrugge2016convergent}.} 

\three{MBS does not suffer from these limitations. Like PSTD, it only requires the spatial sampling to be fine enough to sample the spatial frequencies in the simulated field, i.e., the Nyquist sampling limit for the simulation, allowing large simulation domains to fit in memory. In addition, since the time derivative is also not approximated, MBS requires far fewer (and approximately equally fast) iterations than PSTD. Where PSTD needs 10 or more steps per wavelength, MBS covers the simulation domain in steps of multiple wavelengths at once, a phenomenon known as pseudo-propagation \cite{osnabrugge2016convergent}.} 

\three{These combined benefits currently make the MBS one of the most efficient methods to solve Helmholtz and Maxwell's equations \cite{osnabrugge2016convergent, kruger2017solution, lee2022inverse}, with subsequent extensions to birefringent media \cite{vettenburg2019calculating}, the inverse scattering problem \cite{lee2022inverse}, and applications to sound and seismic waves \cite{huang2019applicability, kaushik2020convergent, stanziola2025iterative}. 
In experiments, the MBS has been demonstrated to be more than three orders of magnitude faster than FDTD \cite{kruger2017solution} and PSTD \cite{osnabrugge2016convergent}.}

\two{A significant speed increase can be achieved by executing the simulations on a graphics processing unit (GPU). For example, Osnabrugge et al. \cite{osnabrugge2021ultra} used MBS on a GPU to solve Maxwell's equations in a structure of $19.7 \cdot 10^3$ cubic wavelengths (sampled at $5$ points per wavelength) in just 3 seconds, representing a factor of $50$ speed-up compared to performing the simulations on a central processing unit (CPU) (in both cases, the Mie sphere simulations were run until the maximum accuracy, a relative error of $0.014$ with the analytical solution, was reached)}. Recently, Valantinas and Vettenburg \cite{valantinas2024scaling} solved the Helmholtz equation for a system of $2.1 \cdot 10^7$ cubic wavelengths \two{sampled at $3$ points per wavelength, reaching a residue of $10^{-4}$ in $13$ minutes and $4270$ iterations}, by mapping the steps of the MBS to layers in an artificial neural network framework, leveraging \one{the memory efficiency of the MBS and} the extensive software and hardware acceleration available for these applications. \one{For comparison, one of the largest recent FDTD simulations was a much smaller $4.6 \cdot 10^5$ cubic wavelengths (sampled at $18$ points per wavelength), which was run for $6235$ time steps on an elastic network of an unspecified number of NVIDIA GPUs and took $5$ minutes \cite{hughes2021perspective}.} \two{However, while GPUs offer significantly faster performances than CPUs, they have a limited amount of working memory.} They cannot be expanded as flexibly as the main working memory, adding a severe limitation to the simulation size \cite{papp2021application, cooper2024speed}. For FDTD, this problem of limited memory has been solved by domain decomposition, i.e., by distributing the computations of a huge problem over multiple PCs or GPUs, each solving a subdomain of the original problem \cite{benamou1997domain, collino2000domain, gander2002optimized, toselli2004domain, bercovier2009domain, dolean2015introduction}. 

\one{Here, we present a method to parallelise the MBS by distributing the computations over multiple GPUs. Our non-overlapping domain decomposition strategy \cite{benamou1997domain, collino2000domain, gander2002optimized, toselli2004domain, bercovier2009domain} is derived directly from a block operator decomposition of the Helmholtz equation. This eliminates the need for the commonly used explicit subdomain boundary conditions (e.g.,  \cite{toselli2004domain, berenger1994perfectly, dolean2015introduction}), and makes it possible to maintain the key benefits of the MBS: low memory use, high accuracy, and guaranteed monotonic convergence, while minimising the overhead on memory use and computation time. We show that the rate of convergence does not deteriorate when the number of subdomains grows, making our method scalable according to the definition in Refs.~\cite{toselli2004domain, bercovier2009domain}.} 

We demonstrate the domain decomposition approach on a GPU system through a large 3D simulation of $3.27\cdot 10^7$ cubic wavelengths \two{sampled at $4$ points per wavelength}, a size that is  $1.95 \times$ larger than the maximum that could be accommodated on a single GPU, and takes only $45$ minutes to \two{reach a relative residual of $10^{-6}$}. \two{With this new framework, a larger simulation size is possible as it is no longer limited to the memory size of a single graphics card. Further, the computations can be parallelised and scaled to even larger domains by adding more GPUs}. An open-source implementation in Python is available on GitHub \cite{code_wavesim_python}. 

We start in Section~\ref{sec:mbs} by summarising the MBS approach. In Section~\ref{sec:domain_decomp}, we introduce the proposed domain decomposition method \one{and discuss its implementation, computational complexity and memory usage}. We then demonstrate the approach and analyse its efficiency and accuracy (Section~\ref{sec:results}). Concluding remarks are provided in Section~\ref{sec:conclusions}.

\section{Modified Born series}
\label{sec:mbs}
\noindent This section summarises the MBS approach introduced by Osnabrugge et al. \cite{osnabrugge2016convergent}, using the simplified formalism introduced in \cite{vettenburg2023universal}. \three{Consider the inhomogeneous Helmholtz equation
\begin{equation}
    \left( \nabla^2 + {k}^2 \right) \psi = -S,
    \label{eq:helmholtz}
\end{equation}
with the Laplacian $\nabla^2$, spatially varying wavenumber ${k}$, field $\psi$, and source $S$. For convenience, we introduce the operator notation 
\begin{equation}
\label{eq:LV}
    A \bm{x} = \bm{y},
\end{equation}
where $A\coloneqq c\left(\nabla^2+k^2\right)$ is a linear operator, $\bm{x}\coloneqq\psi$ and $\bm{y}\coloneqq -c S$ are the field and scaled source sampled on a regular grid, and $c$ is a scaling factor discussed below.}

The key step in both the modified \cite{osnabrugge2016convergent} and original \cite{vanrossum1999multiple} Born series formalisms is to split $A$ into $L+V$, with $L \coloneqq c \left[\nabla^2 + k_0^2 \right]$ corresponding to wave propagation in a homogeneous medium with background wavenumber $k_0$, and $V = c \left[{k}^2 - k_0^2 \right]$ the scattering potential. 

The MBS method uses this splitting to define a preconditioned Richardson iteration \cite{vettenburg2023universal}
\begin{equation}
    \bm{x}^{(n+1)} = \bm{x}^{(n)} + \alpha \Gamma^{-1}\left(\bm{y} - A\bm{x}^{(n)}\right),\label{eq:richardson}
\end{equation}
where $n$ is the iteration counter and $\Gamma = (L+I)(I-V)^{-1}$ is the preconditioner. The constant $\alpha$ is called the relaxation parameter \cite{young1971iterative, samarskij1989numerical, bialecki1994preconditioned}.

In contrast to the original Born series, which diverges for large or strongly scattering media \cite{kleinman1990convergent}, the MBS converges monotonically to the solution of Eq.~\eqref{eq:helmholtz} provided that $\alpha\leq 1$ and $ic$ is a real number less than \two{$1/\norm{k^2 - k_0^2}_\infty$} \cite{osnabrugge2016convergent}, where \two{$\norm{\cdot}_\infty$} denotes the maximum value over all $\bm{r}$. We truncate this iteration when the norm of the residual $\Gamma^{-1}\left(\bm{y} - A\bm{x}^{(n)}\right)$ drops below a certain threshold.

\begin{figure}
    \centering
    \includegraphics[width=\linewidth]{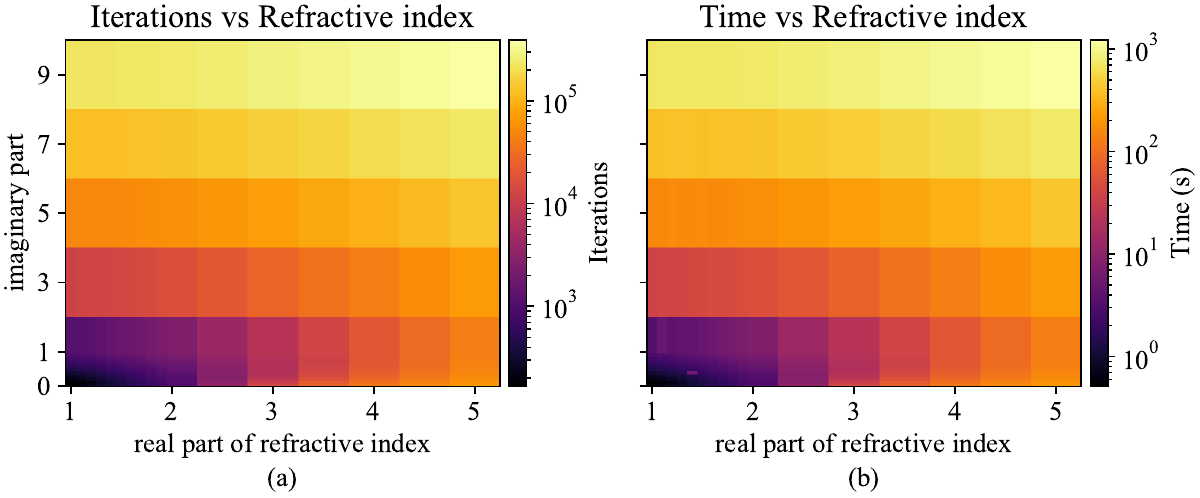}
    \caption{\two{Convergence of the MBS for a problem of size $50\times50\times50$ wavelengths sampled at $4$ points per wavelength (details in Sections~\ref{sec:results} and \ref{subsec:results_validation}) for different refractive index values of the scattering medium (the background refractive index is kept $1$). The real part ranges from $1$ to $5$, and the imaginary part (extinction coefficient) from $0$ to $9$. The (a) number of iterations and (b) simulation time to converge to a relative residual of $10^{-6}$ increase with scattering contrast as expected.}}
    \label{fig:ri_validation}
\end{figure}

A further refinement to optimise the convergence rate was given in \cite{vettenburg2023universal}, leading to the choices 
\begin{equation}\label{eq:c_scale}
    c = -\frac{0.95i}{\norm{{k}^2 - k_0^2}_\infty},\qquad \alpha=0.75,
\end{equation}
\two{which shows that $c$ is inversely proportional to the scattering contrast in the simulation. The number of iterations of the MBS is, in turn, inversely proportional to the scaling factor $c$, so the number of iterations increases with the scattering contrast \cite{osnabrugge2016convergent}. Figure~\ref{fig:ri_validation} demonstrates this effect for a simulation of scalar wave propagation in a collection of spheres (details in Sections~\ref{sec:results} and \ref{subsec:results_validation}). The refractive index (or relative wave velocity) of the scattering medium ranges from $1$ to $5$ for the real part and from $0$ to $9$ for the imaginary part, while the background refractive index is $1$. It can be seen from the figure that the MBS converges for all refractive index values considered, and the number of iterations (and simulation time) increases with the scattering contrast.}

\begin{figure}
\centering\includegraphics[width=\linewidth]{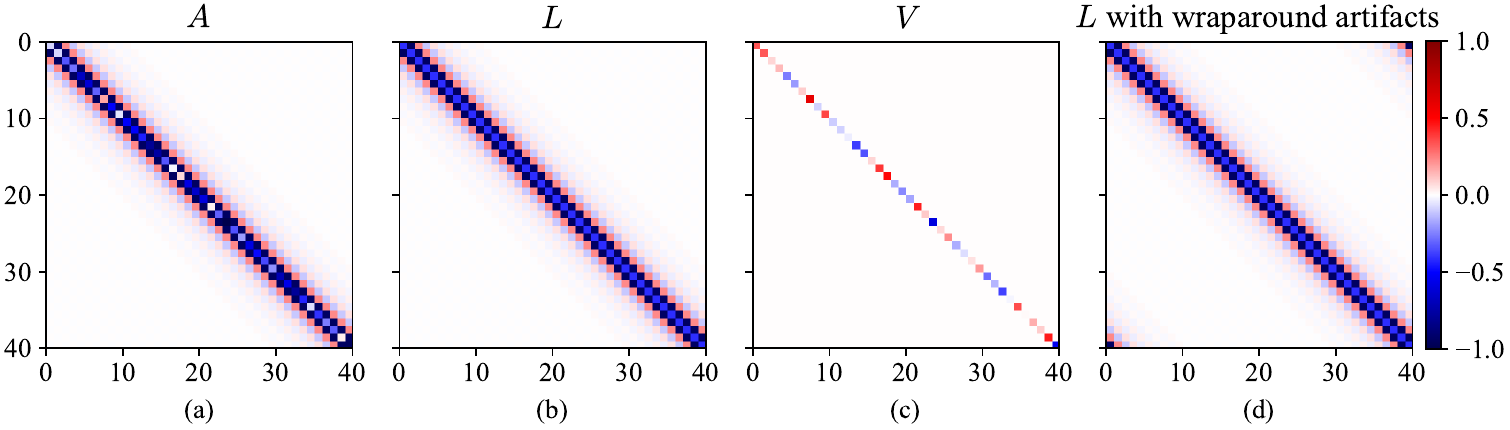}
\caption{Matrix representation of splitting operator $A$ into $L + V$ for a 1D system. (a) Operator $A$ is split into (b) $L$ containing the Laplacian operator, a band along the diagonal, and (c) $V$, a diagonal matrix containing the scattering potential. (d) The operator $L$ with wraparound artefacts resulting from implementing $L$ as a fast convolution (Eq.~\eqref{eq:L-convolution}).}
\label{fig:splitting}
\end{figure}

The splitting $A=L+V$ is visualised for a 1D problem in Fig.~\ref{fig:splitting}a-c, where the operators are represented as matrices for illustrative purposes. In practice, these operators are not implemented as matrices, as the full matrices would be too large to store in memory. The alternative of using sparse matrices would require replacing $L$ by a matrix with only a few diagonals, which is precisely the finite difference approximation that causes numerical dispersion and thus should be avoided. Instead, $L$ can be implemented as a convolution with the angular spectrum kernel \cite{sherman1969diffracted, oughstun1998angular}
\begin{equation}\label{eq:L-convolution}
    L = c\inv{\mathcal{F}} \left( -\norm{\bm{p}}_2^2 + k_0^2 \right) \mathcal{F},
\end{equation}
where $\mathcal{F}$ and $\inv{\mathcal{F}}$ denote the forward and inverse Fourier transform, respectively, $\bm{p}$ denotes the Fourier-space coordinate vector, \two{and $\norm{\cdot}_2$ is the Euclidean norm}. In practice, a fast Fourier transform (FFT) is used to efficiently evaluate this convolution. As shown in Fig.~\ref{fig:splitting}d, this approach does introduce wrapping artefacts. These artefacts can be significantly reduced by using an alternating sequence of offset Fourier transforms \cite{osnabrugge2021ultra} and eliminated using our proposed method (see Section~\ref{sec:domain_decomp}). \three{Either of these methods can then be combined with a thin anti-reflection layer \cite{osnabrugge2021ultra} to prevent reflections at the domain boundaries}.

Similar to Eq.~\eqref{eq:L-convolution}, we can implement $(L+I)^{-1}$ as a (fast) convolution. Substituting $\Gamma^{-1}=(I-V)(L+I)^{-1}$ and $A=(L+I) - (I-V)$ into Eq.~\eqref{eq:richardson} gives the iteration \cite{osnabrugge2016convergent}
\begin{equation}\label{eq:wavesim_iter}
    \bm{x}^{(n+1)} = \bm{x}^{(n)} - \alpha (I-V)\left(\bm{x}^{(n)}-(L+I)^{-1}\left[\bm{y} + (\two{I}-V)\bm{x}^{(n)}\right]\right),
\end{equation}
which requires only one fast convolution per iteration. This is another important benefit of the MBS: the preconditioned operator $\Gamma^{-1}A$ can be evaluated at almost no extra cost compared to the original operator $A$, which also requires a single fast convolution.

\section{Domain decomposition of the modified Born series}
\label{sec:domain_decomp}
\begin{figure}
\centering\includegraphics[width=0.76\linewidth]{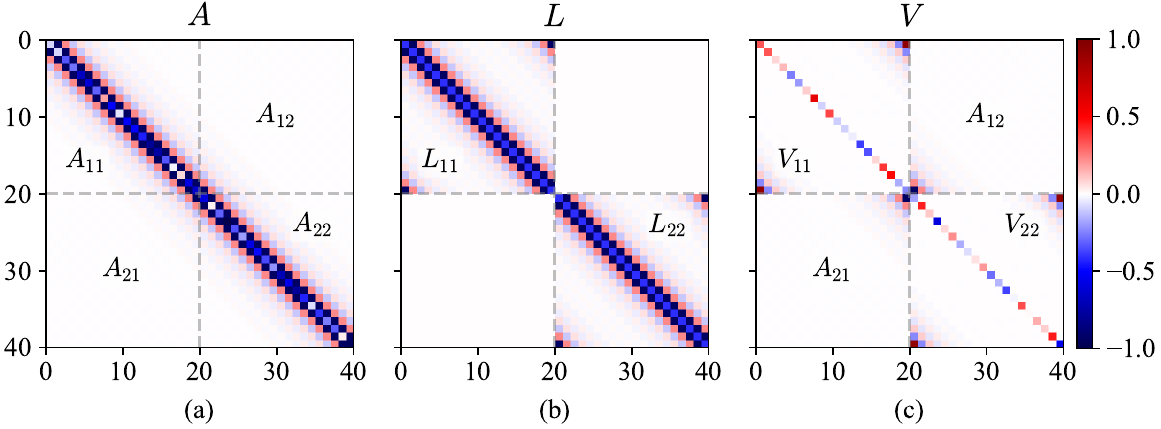}
\caption{Matrix representation of the block decomposition of Eq.~\eqref{eq:LV}. (a) $A$ (same as Fig.~\ref{fig:splitting}a) is decomposed over two subdomains, shown by the dashed grey lines. (b) Decomposition of $L$, implemented as a fast convolution (Eq.~\eqref{eq:L-convolution}) over each subdomain, thus containing wraparound artefacts. \one{The off-diagonal (communication) blocks are empty. (c) $V \coloneqq A - L$ has diagonal blocks containing the scattering potential on the diagonal elements and wrapping artefacts from $L$ but with the opposite sign, and off-diagonal blocks representing communication between the subdomains.}}
\label{fig:decompose}
\end{figure}

\noindent \one{To parallelise the MBS so that it can be executed on multiple GPUs simultaneously,} we start by decomposing the linear system $A \bm{x} = \bm{y}$ \one{(Eq.~\eqref{eq:LV})} over multiple subdomains. For illustrative purposes, we decompose a 1D problem into two subdomains of equal size $N$. The same strategy extends to 2D and 3D problems, with any number of subdomains along any dimension and domains of unequal size. The operator $A$ for this 1D problem can be represented as a block \one{decomposition}:

\begin{align}
    \begin{bmatrix}
        A_{11} & A_{12} \\
        A_{21} & A_{22}
    \end{bmatrix} 
    \begin{bmatrix}
        \bm{x}_{1} \\
        \bm{x}_{2}
    \end{bmatrix}
    &= 
    \begin{bmatrix}
        \bm{y}_{1} \\
        \bm{y}_{2}
    \end{bmatrix},
    \label{eq:A_block}
\end{align}
where the blocks $A_{11}, A_{12}, A_{21}$, and $A_{22}$ are of size ${N \times N}$, and the vectors $\bm{x}_1, \bm{x}_2, \bm{y}_1$, and $\bm{y}_2$ are of length $N$. \one{Here, the blocks $A_{11}$ and $A_{22}$ operate on subdomains 1 and 2, respectively, while the blocks $A_{12}$ and $A_{21}$ represent communication between the subdomains. The decomposition of $A$ over two subdomains is visualised in Fig.~\ref{fig:decompose}a}. \one{Note that, as before, these blocks are linear operators and not stored as matrices explicitly.}

An essential generalisation of the MBS introduced in \cite{vettenburg2023universal} is that we have complete freedom to choose the splitting $A=L+V$, as long as the scaling factor $c$ can be selected such that $\norm{V}<1$, \two{where $\norm{\cdot}$ denotes the operator norm, i.e., its largest singular value.} We use this freedom to implement our domain decomposition framework.

This freedom enables us to implement $L$ as a fast convolution over each subdomain (see Fig.~\ref{fig:decompose}b)\one{, with $V\coloneqq A - L$ \cite{vettenburg2023universal}}. \one{Representing $L$ and $V$ as a block decomposition
\begin{equation}\label{eq:LV_block}
    L =   \begin{bmatrix}
                L_{11} & 0 \\
                0 & L_{22}
            \end{bmatrix}, \quad
    V =   \begin{bmatrix}
                A_{11} - L_{11} & A_{12} \\
                A_{21} & A_{22} - L_{22}
            \end{bmatrix}
      \coloneqq \begin{bmatrix}
                    V_{11} & A_{12} \\
                    A_{21} & V_{22}
                \end{bmatrix},
\end{equation}
we see that $L$ (visualised in Fig.~\ref{fig:decompose}b) is a domain-local operator, with the diagonal blocks $L_{11}$ and $L_{22}$ operating on the subdomains, and contains wraparound artefacts similar to Fig.~\ref{fig:splitting}d, but now over the subdomains. $V$ now has both diagonal and off-diagonal blocks (visualised in Fig.~\ref{fig:decompose}c). The diagonal blocks $V_{11}$ and $V_{22}$ contain, as before, the scattering potential on the diagonal elements (as in Fig. \ref{fig:splitting}c), and now also the wrapping artefacts from $L_{11}$ and $L_{22}$ but with the opposite sign. The off-diagonal blocks $A_{12}$ and $A_{21}$ represent communication between the subdomains.} 

\one{Unlike common domain decomposition methods \cite{benamou1997domain, collino2000domain, toselli2004domain, dolean2015introduction, gander2019class, modave2020nonoverlapping, royer2022nonoverlapping}, our approach does not require the explicit specification of boundary conditions at the subdomain boundaries, because when added up, $L+V$ is still exactly identical to the original system (Eq.~\eqref{eq:LV}).}

\begin{figure}
        \centering\includegraphics[width=\linewidth]{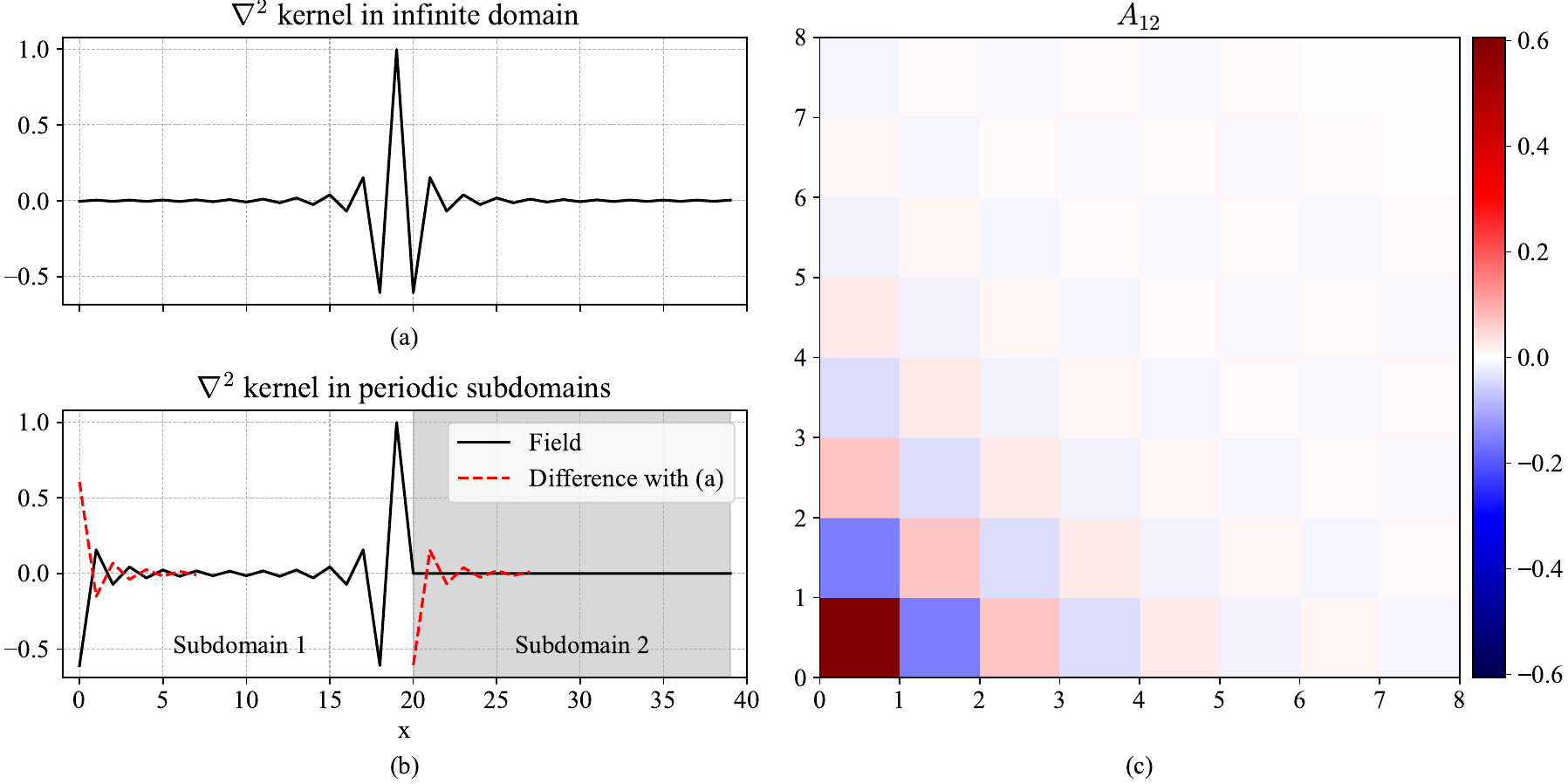}
\caption{Generation of the off-diagonal blocks $A_{12}$ and $A_{21}$ using the angular spectrum kernel. $A_{21}=A_{12}^T$, so we only show $A_{12}$. (a) Field after convolution of the Laplacian with a point source. (b) Field when the fast convolution is computed only over Subdomain 1 (white background). The difference between the fields in (a) and (b) is shown by the dashed red lines and gives columns of (c) the (truncated) block $A_{12}$.}
\label{fig:correction_matrix}
\end{figure}

\one{In our framework, we do need a way to efficiently compute the off-diagonal elements in $V$. We realise that these elements represent the difference between the kernel applied to a point source in an infinite domain (Fig.~\ref{fig:correction_matrix}a) versus a periodic subdomain (Fig.~\ref{fig:correction_matrix}b). Fig.~\ref{fig:correction_matrix}c visualises this process repeated for different positions of the point source, giving the bottom-left part of $A_{12}$ (also see Fig.~\ref{fig:decompose}a and c). It can be seen that the magnitude of the elements decreases rapidly with increasing distance from the boundary. Therefore, we can truncate the off-diagonal blocks and keep only $t\ll N$ points, so that $A_{12}$ can be implemented as a $t\times t$ matrix-vector multiplication using only the first or last $t$ points of each edge of the subdomain.}

\one{Due to the symmetry of the angular spectrum kernel, the off-diagonal elements of $V_{11}$ ($=V_{22}$) are equal to $-(A_{12}+A_{12}^T)$ (see Fig.~\ref{fig:decompose}d). In other words, we can use the same operator (up to a minus sign) to compute both the transfer between subdomains and the wrapping correction inside each subdomain.}

\one{Note that this approach also eliminates the wrapping artefacts present in the original system (the corners in Fig.~\ref{fig:splitting}d). We therefore eliminate the need for expensive zero padding or perfectly matched layers \cite{berenger1994perfectly, liu1997pstd}, and instead use a thin anti-reflective boundary \cite{osnabrugge2021ultra}, reducing the computational costs and memory requirements.}

\begin{figure}
\centering\includegraphics[width=0.76\linewidth]{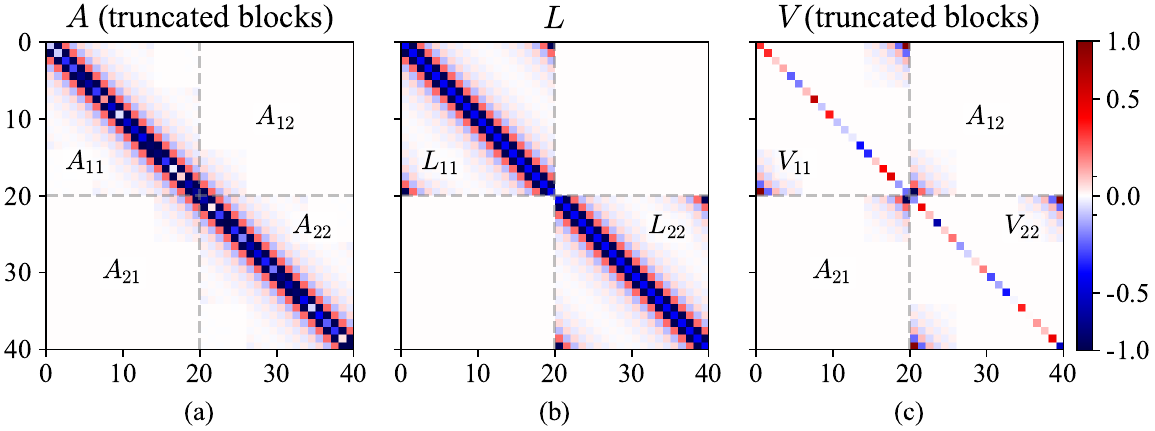}
\caption{\one{Using truncated blocks in $V$ to reconstruct $A$. (a) Matrix representation of $A$ decomposed over two subdomains (demarcated by the dashed grey lines), but with truncated blocks. (b) $L$ decomposed over the subdomains remains unchanged and is the same as in Fig.\ref{fig:decompose}b. (c) $V$ with truncated blocks for communication (off-diagonal) and removing wrapping artefacts (off-diagonal elements of the diagonal blocks), visible as square blocks of side $t=6$ in the corners of subdomain blocks. Note that even though the colour scale is logarithmic in both the positive and negative directions from 0, the blocks are hardly visible, showing that the effect of the truncation is small.}}
\label{fig:truncate}
\end{figure}

\one{Figure~\ref{fig:truncate} is a reproduction of Fig.~\ref{fig:decompose} but with truncated blocks for communication between subdomains (the off-diagonal blocks $A_{12}$ and $A_{21}$) and for removing the wrapping artefacts (the off-diagonal elements of $V_{11}$ and $V_{22}$). The operator $A$ obtained using these truncated blocks is shown in Fig.~\ref{fig:truncate}a.} The decomposed operator $L$ (Eq.~\eqref{eq:LV_block}) shown in Fig.~\ref{fig:truncate}b remains unchanged irrespective of the truncation and is the same as in Fig.~\ref{fig:decompose}b. \one{Figure~\ref{fig:truncate}c shows $V$ obtained using the truncation instead of the full blocks.} \two{The effect of the truncation is hardly visible in Fig.~\ref{fig:truncate}c as square blocks of size $t=6$, indicating that the effect of the truncation is small. In Section~\ref{subsubsec:results_truncation}, we analyse quantitatively the effect of the truncation parameter $t$ on the accuracy of the simulation.}

\subsection{Implementation}
\label{subsec:implementation}
\noindent The domain decomposition framework is incorporated into the preconditioned Richardson iteration (Eq.~\eqref{eq:wavesim_iter}), but with the operators $L$ and $V$ computed for and applied to each subdomain. Thus, the fast convolution with $\inv{(L+I)}$ is computed for each subdomain $s$ as $\inv{(L_{ss}+I)}$ (see Figs.~\ref{fig:decompose}b and \ref{fig:truncate}b). 

\one{As discussed in Section~\ref{sec:domain_decomp}, operator $V$ has three contributions, i.e., the scattering potential on the diagonal, the communication between the subdomains (off-diagonal blocks) and the wrapping artefacts with the opposite sign to $L$ in the diagonal blocks. The steps for applying the operator $V_{ss}$ for every subdomain $s$ are as below:
\begin{enumerate}
    \item Compute the edge corrections (e.g., $A_{12}\bm{x}_{2}$) for all faces of all subdomains, using only $t$ voxels at the edges.
    \item Transfer the edge corrections to neighbouring subdomains.
    \item Apply the scattering potential, i.e., the diagonal elements of $V_{ss}$.
    \item \emph{Subtract} the edge corrections from the current subdomain (removes wrapping artefacts).
    \item \emph{Add} the edge corrections coming from the neighbouring subdomains (adds the field entering from the neighbouring subdomains).
\end{enumerate}}

\one{The above procedure replaces the original implementation of $V$ and does not affect the iteration in Eq.~\eqref{eq:wavesim_iter} itself.}

\two{We use the relative residual $\epsilon\coloneqq\norm{\Gamma^{-1} \left( \bm{y} - A\bm{x}^{(n)} \right)}_2/\norm{\inv{\Gamma} \bm{y}}_2$ as a stopping criterion for the iteration}. For the domain decomposition case, the global residual is given by $\sqrt{\sum_s \epsilon_s^2}$, where the summation is over all subdomains $s$. We use a threshold value of $10^{-6}$ as the stopping criterion. After the iteration has converged, the fields from the different subdomains are combined to give the final solution of the global field $\bm{x}$.

As we now use a different operator $V$ than the original MBS implementation, we can no longer use Eq.~\eqref{eq:c_scale} to compute the scale factor $c$. \two{We need to recompute $\norm{V}$, the operator norm of $V$, for the domain decomposition case.} For simplicity, we estimate this norm as the sum of the norms of all contributions to $V$: the original scattering potential, \one{the subdomain communication blocks and the wrapping artefacts removal blocks}, which gives an upper limit for $\norm{V}$ and results in a scaling factor of
\begin{equation}\label{eq:c_scale2}
    c = -\frac{0.95i}{\norm{{k}^2 - k_0^2}_\infty + \left(\sum_{d=1}^3 a_d\right)\norm{A_{12}}},
\end{equation}
\one{where $a_d$ equals $0$ if no subdomain communication or removal of wrapping artefacts is applied along dimension $d$ (corresponding to periodic boundaries). It equals $1$ if only wrapping artefacts are removed (i.e., when there is only a single subdomain along this dimension and thus no subdomain communication) and $2$ if both the subdomain communication and removal of wrapping artefacts are applied. Thus, the subdomain communication and removal of wrapping artefacts reduces the value of $c$, leading to slower convergence when $a_d \neq 0$. It is important to note, however, that $a_d$ is independent of the number of domains along a dimension. So, increasing the number of subdomains along a dimension $d$ has no additional cost in terms of iterations. We discuss this in detail in Section~\ref{subsubsec:results_subdomains}}. For a 1-domain simulation without removal of wrapping artefacts, $a_d=0$, and Eq.~\eqref{eq:c_scale2} reduces to Eq.~\eqref{eq:c_scale}. 

Domain decomposition also offers an opportunity to improve the efficiency of the computations by initially activating only those subdomains that hold a source term. Neighbouring subdomains are activated only after the norm of the transferred field exceeds a certain threshold. Although this initially causes a small error, the MBS converges to the correct solution from any starting point and thus nullifies these errors. \two{The number of iterations does not change due to this activation strategy, but the simulation time reduces, with the gains increasing with the number of domains and the simulation size.} For example, a 3D simulation of size $100 \times 100 \times 100$ wavelengths split into $3 \times 1 \times 1$ subdomains was $12\%$ faster than without this strategy.

\subsection{\two{Complexity analysis}}
\label{subsec:complexity_analysis}
\two{Table~\ref{tab:compute_complexity} compares the parameters of our domain decomposition method with the original MBS \cite{osnabrugge2016convergent}, a recent FFT-accelerated solver for the wave equation \cite{vico2016fast}, PSTD \cite{liu1997pstd}, and FDTD methods \cite{yee1966numerical, taflove1995computational}.}

\two{The accuracy of our domain decomposition approach, the original MBS, and the FFT-accelerated solver by Vico et al. \cite{vico2016fast} is only restricted by machine precision, whereas that of FDTD is proportional to $\Delta x^2$ and $\Delta t^2$, and that of PSTD is proportional to $\Delta t^2$ due to the finite different approximations. Although our domain decomposition method and the MBS have no time dependence, we can interpret the `pseudo-propagation' length \cite{osnabrugge2016convergent}, i.e. the distance the solution expands with each iteration, to the real propagation steps of the PSTD and FDTD methods. The pseudo-propagation distance, given by $\lvert 2k_0c\rvert$, depends on the scaling parameter $c$ for both the original MBS (see Eq.~\eqref{eq:c_scale}) and our domain decomposition method (see Eq.~\eqref{eq:c_scale2}).}

\two{The computational complexity of the original MBS implementation has been studied in literature \cite{osnabrugge2016convergent, kruger2017solution}, and it is dominated by the computation of the FFT in the MBS algorithm. For a $D$-dimensional problem with $N^D$ voxels, the method requires $\mathcal{O}\left(N^D\textrm{log}N^D\right)$ floating point operations per iteration, the same as the FFT-accelerated solver by Vico et al. \cite{vico2016fast} and the PSTD method \cite{liu1997pstd}.} 

\two{For the domain decomposition method, we divide the $N^D$ voxels over $M$ total subdomains. Then we need to compute a total of $DM$ FFTs, each over a subdomain of size $N^D/M$, giving a complexity of $\mathcal{O}\left(N^D\textrm{log}(N^D/M)\right)$ for the FFTs. The overhead due to the domain decomposition is $\mathcal{O}\left(MDN^{D-1}t\right)$, which is asymptotically smaller than that of the FFT.}

\begin{table}
    \centering
    \small
    \begin{tabular}{L{1.9cm}|C{2.5cm}|C{0.8cm}|C{1.4cm}|C{2.5cm}|C{1.9cm}}
         & \two{This work} & \two{MBS \cite{osnabrugge2016convergent}} & \two{Vico et al. \cite{vico2016fast}} & \two{PSTD \cite{liu1997pstd}} & \two{FDTD \cite{yee1966numerical, taflove1995computational}} \\
        \hline
        \hline
        \two{Grid spacing $\Delta x$} & \multicolumn{4}{c|}{\two{$\leq \lambda_m/2$}} & \two{$\leq \lambda_m/8$} \\
        \hline
        \multirow{2}{*}{\two{Accuracy}} & \multicolumn{3}{c|}{\two{tolerance adjustable}} & \multirow{2}{*}{\two{$\mathcal{O}(\Delta t^2)$}} & \multirow{2}{*}{\two{$\mathcal{O}(\Delta t^2\Delta x^2)$}} \\
        & \multicolumn{3}{c|}{\two{to machine precision}} & & \\
        \hline
        \two{Time step $\Delta t$} & \multicolumn{3}{c|}{\two{n/a}} & \two{$\leq \frac{2\Delta x}{\pi v_c\sqrt{D}}$} & \two{$\leq \frac{\Delta x}{v_c\sqrt{D}}$} \\
        \hline
        \two{(Pseudo-) propagation} & \multicolumn{2}{c|}{\two{$\lvert 2k_0c\rvert$}} & \two{\footnotesize no estimate available} & \two{$\frac{2}{k_m\sqrt{D}}$} & \two{$\frac{\pi}{4k_m\sqrt{D}}$} \\
        \hline
        \two{FLOPs per iteration} & \two{$\mathcal{O}\left(N^D\textrm{log}\frac{N^D}{M}\right)$} & \multicolumn{3}{c|}{\two{$\mathcal{O}(N^D\textrm{log}N^D)$}} & \two{$\mathcal{O}(N^D)$}\\
        \hline
        \two{Number of iterations} & \multicolumn{2}{c|}{\two{$\mathcal{O}\left(\frac{Tv_c}{\lvert k_0c \rvert}\right)$}} & \two{\footnotesize no estimate available} & \two{$\mathcal{O}(Tv_ck_m \sqrt{D})$} & \two{$\frac{T}{\Delta t}$} \\
        \hline
    \end{tabular}
    \caption{\two{Key parameters for our domain decomposition approach (``This work''), compared to the original MBS \cite{osnabrugge2016convergent}, the fast Fourier transform (FFT)-accelerated solver in \cite{vico2016fast}, the PSTD \cite{liu1997pstd} and FDTD \cite{yee1966numerical, taflove1995computational} methods, following Table 1 from Osnabrugge et al. \cite{osnabrugge2016convergent}. Here, $k_m\coloneqq \norm{k^2}_\infty$ is the largest wavenumber, $\lambda_m\coloneqq2\pi/k_m$ is the shortest wavelength, $D$ is the dimensionality of the system, $v_c$ is the speed of light in the simulated medium, and $T$ is the time needed for a wave to propagate through the full simulation.}}
    \label{tab:compute_complexity}
\end{table}

\subsection{\two{Memory usage}}
\label{subsec:memory}
\two{The memory usage of the MBS method is approximately $40$ bytes per voxel. The factor $40$ originates from $5$ arrays made up of complex floats of $8$ bytes each, namely, the refractive index, the ${(L+I)}^{-1}$ operator, the $(I-V)$ operator (Eq.~\eqref{eq:wavesim_iter}), an array for updating and storing the field $\bm{x}$, and an internal FFT array.}

\two{On top of this, our domain decomposition strategy requires approximately $8 \cdot 2 \cdot t$ bytes per voxel at each subdomain interface, with $8$ bytes per complex float and $t$ the truncation parameter. We use a default value of $t=8$ in our simulations, so the relation reduces to $8 \cdot 2 \cdot 8 = 128$ bytes per voxel at each subdomain interface.}

\two{For instance, we can compute the memory usage for the $320 \times 320 \times 320$ wavelengths structure sampled at $4$ points per wavelength that we later present results for in Section~\ref{subsec:results_large3D}. We use absorbing boundaries of $5$ wavelengths along the x-axis and periodic boundaries along the y and z axes, giving the resultant simulation size of $330 \times 320 \times 320$ wavelengths, or $1320 \times 1280 \times 1280$ voxels ($2.16\times10^9$ voxels). The memory needed for a one-domain simulation is $40 \cdot 1320 \cdot 1280^2 = 86.5$ GB. The largest VRAM with a single GPU available to us is $48$ GB with the NVIDIA A40 GPUs, so we split the problem over two GPUs into two domains along the x-axis, and there are $1280^2$ voxels at the subdomain interface. The overhead is calculated as only $128 \cdot 1280^2 = 0.2$ GB, giving a total memory use of $86.7$ GB for the largest simulation we could run on two GPUs with a total memory of $96$ GB, indicating an unaccounted overhead of around $9.7\%$ in the memory usage calculations.}

\section{Results and Discussion}
\label{sec:results}
\noindent In this section, we demonstrate the domain decomposition of the MBS in 3D simulations. For all simulations in this section, we simulate light propagation through a scattering structure illuminated with a planar source located at $x=0$. We use absorbing boundaries \cite{osnabrugge2021ultra} with a thickness of 5 wavelengths along the x-axis and periodic boundaries along the y and z axes. We use a sampling of 4 points per wavelength\two{, and a threshold value of $10^{-6}$ for the relative residual as the stopping criterion (see Section~\ref{subsec:implementation}).} \one{In all the simulations, we split the problem into subdomains of equal size.}

The structure consists of tightly packed spheres, with a refractive index of $1.33$ and a small imaginary part \two{of $0.01$}, distributed randomly in a medium with a refractive index of $1$. The spheres were generated using the \texttt{random\_spheres()} function in PoreSpy \cite{gostick2019porespy}. Unless stated otherwise, the spheres have a radius of $3$ wavelengths and the truncation parameter $t=8$ in the simulations.

We first confirmed that our domain decomposition method converges to the analytical solution for propagation through an empty medium. In this section, we start by validating the domain decomposition framework on a small sample and then analyse the accuracy and convergence of the domain decomposition approach for different values of the truncation parameter $t$ and numbers of subdomains. Finally, we demonstrate domain decomposition in a large structure of $3.27\cdot 10^7$ cubic wavelengths\two{, a size that takes considerably longer to run on a CPU but can be completed significantly faster by splitting the problem over two GPUs}.

\subsection{Domain decomposition validation}
\label{subsec:results_validation}
\noindent We first validate the accuracy of our domain decomposition framework by comparing it to the single-domain MBS approach for a 3D problem of size $50 \times 50 \times 50$ wavelengths. \one{With absorbing boundaries \cite{osnabrugge2021ultra} of $5$ wavelengths along the x-axis, the total simulation size is $60 \times 50 \times 50$ wavelengths}. We quantify the accuracy by comparing the field obtained from domain decomposition ($\bm{x}$) and the single-domain simulation result ($\bm{x}_{\mathrm{ref}}$), excluding the absorbing boundaries, with the relative error $\norm{\bm{x} - \bm{x}_{\mathrm{ref}}}_2^2/\norm{\bm{x}_{\mathrm{ref}}}_2^2$.

\begin{figure}
    \centering
    \includegraphics[width=\linewidth]{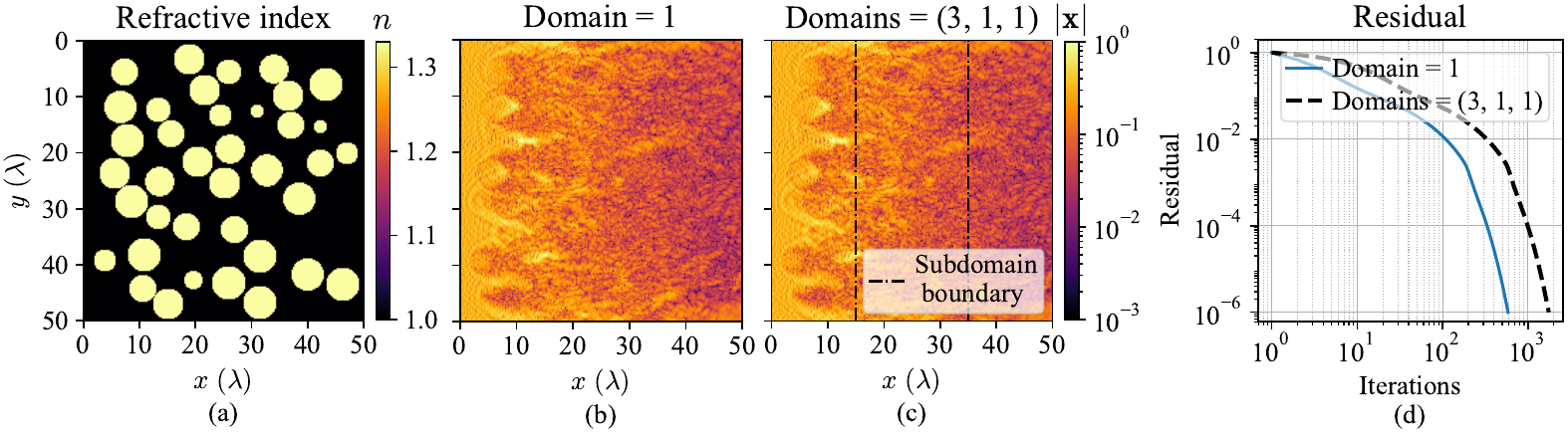}
    \caption{Simulation results with and without domain decomposition for a 3D medium of size $50 \times 50 \times 50$ wavelengths. 2D slices at z/2 of (a) the refractive index distribution used in the simulation, (b) the result for a single domain, and (c) the result with three subdomains in the x direction, with the dash-dotted black lines indicating the subdomain boundaries. \two{The spheres appear to be of different sizes because the 2D slice takes a cross-section of the spheres at different points.} (d) Residual vs. iteration number for the simulations in b (solid blue line) and c (dashed black line).}
    \label{fig:results_validation}
\end{figure}

Fig.~\ref{fig:results_validation}a shows a slice through the centre of the z-axis of the 3D refractive index distribution used in this test, without absorbing boundaries. The results of the simulations are shown without absorbing boundaries in  Fig.~\ref{fig:results_validation}b for the single domain \one{of simulation size $60 \times 50 \times 50$ wavelengths} and in Fig.~\ref{fig:results_validation}c for the domain decomposed into three subdomains along the x-axis\one{, each of size $20\times50\times50$ wavelengths}. The relative error between the two is just $2\cdot 10^{-4}$, confirming the validity of our domain decomposition approach.

Fig.~\ref{fig:results_validation}d shows the residual as a function of the iteration number for the two simulations. It can be seen that the residual decreases monotonically, which is a key property of the MBS and is also conserved in our domain decomposition method. The single-domain simulation converges (i.e., reaches a relative residual of $10^{-6}$) in 584 iterations in 6.7 seconds, while the 3-domain simulation converges in 1751 iterations in 8.1 seconds. The $3 \times$ increase in the number of iterations is due to a lower scaling coefficient $c$, with $a_d=2$ (Eq.~\eqref{eq:c_scale2}). Although this is a significant increase in the number of iterations, this factor is independent of the number of subdomains along an axis, as we analyse in more detail in Section~\ref{subsubsec:results_subdomains}. 

\subsection{Convergence dependence on truncation parameter and number of subdomains}
\label{subsec:results_convergence}
\noindent In this section, we test the convergence behaviour of the domain decomposition simulations and the dependence on the truncation parameter and number of subdomains using the same structure as before.
\subsubsection{Dependence on truncation parameter}
\label{subsubsec:results_truncation}
\begin{figure}
    \centering
    \includegraphics[width=\linewidth]{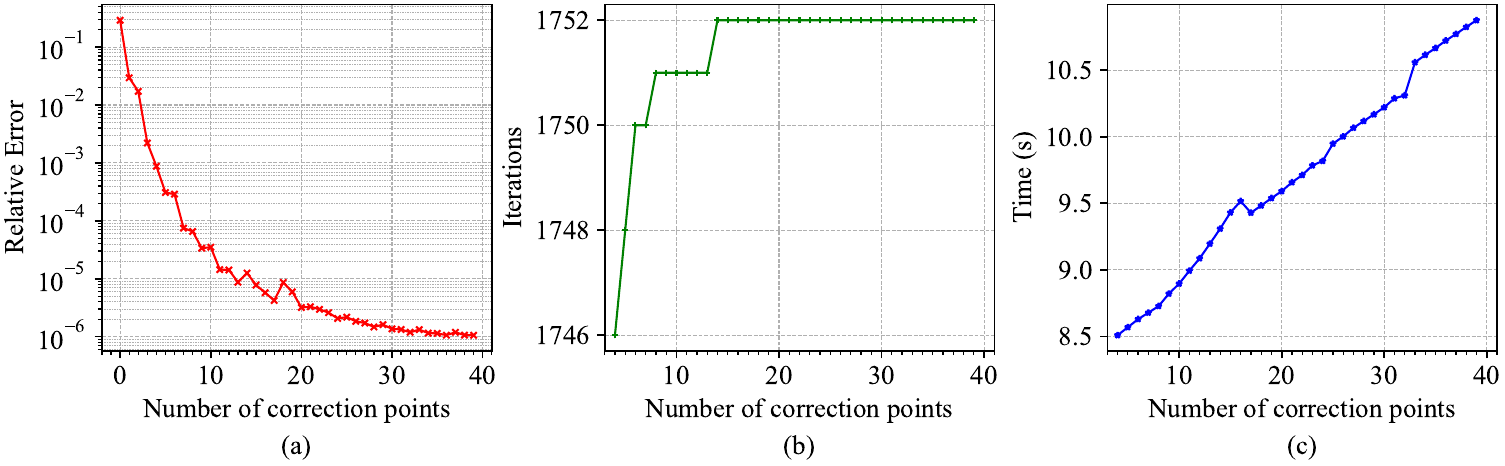}
    \caption{Convergence of a 3D simulation with two subdomains for different values of the truncation parameter $t$. (a) Relative error with a 1-domain simulation decays with $t$. (b) The number of iterations is almost constant for $t=4$ and higher, showing no dependence on $t$. (c) The total simulation time in seconds consistently increases with $t$.}
    \label{fig:dd_truncation}
\end{figure}

\noindent First, we examine the convergence behaviour as a function of the truncation parameter $t$ used to generate \one{the blocks for subdomain communication and removal of wrapping artefacts}. We decompose the problem in Section~\ref{subsec:results_validation} into two subdomains \one{of equal size} along the x-axis \one{($30\times50\times50$ wavelengths each)} and vary $t$ from $0$ to $40$. We compare the relative error of these simulations with a single-domain simulation, as shown in Fig.~\ref{fig:dd_truncation}a. The relative error between the two decays with increasing $t$, and with just $t=4$, the error is already less than $0.1\%$.

Fig.~\ref{fig:dd_truncation}b shows the number of iterations for convergence as a function of $t$, plotted for $t=4$ and higher, as the accuracy is low for $t<4$. The figure shows that the number of iterations does not depend on $t$. The time for convergence (Fig.~\ref{fig:dd_truncation}c) consistently increases with $t$, which is expected as the size of the blocks for subdomain communication and removal of wrapping artefacts increases.
\subsubsection{Dependence on number of subdomains}
\label{subsubsec:results_subdomains}
\begin{figure}
    \centering
    \includegraphics[width=0.76\linewidth]{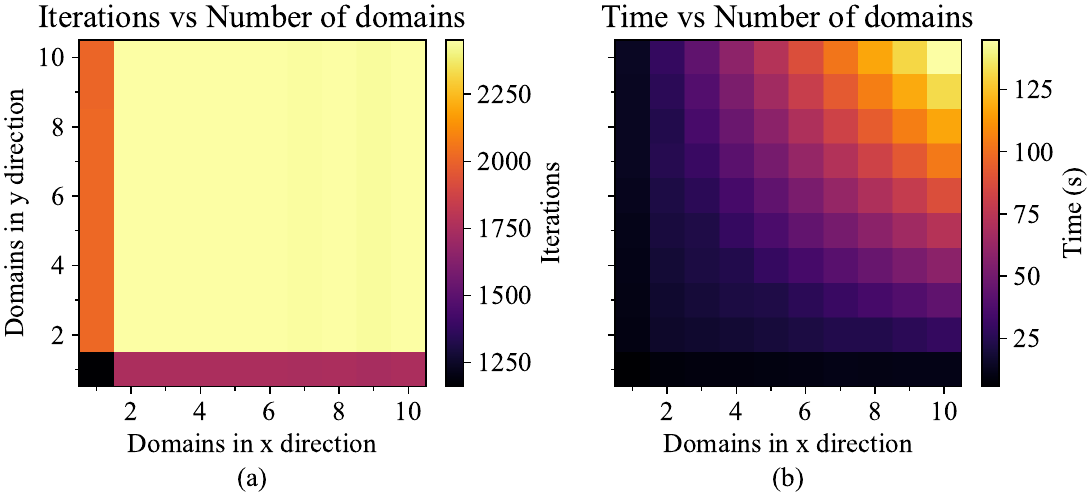}
    \caption{Number of iterations (a) and time (b) as a function of the number of subdomains along the x and y directions for a 3D simulation. The iteration count increases only when subdomains are added along a new axis\two{, which follows from the scaling factor $c$ in Eq.~\eqref{eq:c_scale2} being independent of the number of subdomains.} The time increases with the number of subdomains, but is consistently low for one domain in either x \emph{or} y direction. \two{Note that in both (a) and (b), the colour axis does not start from $0$.}}
    \label{fig:dd_convergence}
\end{figure}

\noindent Next, we examine the number of iterations and time as a function of the number of subdomains (Fig.~\ref{fig:dd_convergence}). Fig~\ref{fig:dd_convergence}a shows the number of iterations for different numbers of subdomains along the x and y-axis. The lowest number of iterations (1160) is for a 1-domain simulation (black pixel in the lower left corner in Fig.~\ref{fig:dd_convergence}a). This case corresponds to a scaling coefficient $c$ based on only the removal of wrapping artefacts along the x-axis, for which the system is non-periodic. When there is more than one subdomain along \emph{either} the x or y-axis, the scaling factor $c$ reduces, which increases the iterations to around $1751$ with more domains along the x-axis (dark pink band of pixels for $y=1$ in Fig.~\ref{fig:dd_convergence}a), and around $2018$ for more domains along the y-axis (orange band of pixels for $x=1$ in Fig.~\ref{fig:dd_convergence}a). For more than one subdomain in \emph{both} the x and y direction, the scaling coefficient $c$ is even lower, increasing the iterations to around $2448$ (Fig.~\ref{fig:dd_convergence}a, the light yellow pixels for $x$ and $y > 1$). Thus, the number of iterations increases only when subdomains are added along a new axis. \two{The number of iterations is not dependent on the number of subdomains beyond one along an axis, indicating the scalability of our domain decomposition framework. This result also follows from the modified scaling factor $c$ in Eq.~\eqref{eq:c_scale2}, which has no dependence on the number of subdomains.}

Fig.~\ref{fig:dd_convergence}b shows the simulation time in seconds as a function of the number of subdomains along the x and y-axis. The simulation time is consistently low for one domain in the x or y directions (black band of pixels along the x and y axes at $x=1$ or $y=1$). It consistently increases with more subdomains due to an increase in the amount of communication between them. The time is highest for $10$ domains along both x and y directions, i.e., the total number of subdomains = $100$ (top right corner of Fig.~\ref{fig:dd_convergence}b).

\subsection{Large scale simulations}
\label{subsec:results_large3D}
\noindent To demonstrate the larger simulation size enabled by the domain decomposition approach, we solve a problem of $3.27\cdot 10^7$ cubic wavelengths \one{($320 \times 320 \times 320$ wavelengths)} on a system with two Silver-4216 2.10 GHz CPUs with 128 GB RAM and four A40 48GB GPUs. The structure is the same as mentioned at the beginning of Section~\ref{sec:results}, except that the spheres have a radius of 6 wavelengths and the clearance between them is set to 6 wavelengths, which helps the energy propagate deeper into the simulation away from the source. \one{The simulation size is $330 \times 320 \times 320$ wavelengths (with absorbing boundaries of $5$ wavelengths along the x-axis), corresponding to $1320 \times 1280 \times 1280$ voxels ($2.16$ Gigavoxels).} Due to this large size, the simulation cannot be performed on a single GPU. \two{It can, however, be run on a CPU, with a significant tradeoff in terms of computation time.} Therefore, we run the single-domain simulation on the CPU and compare it against the 2-domain, 2-GPU simulation performed with the same system. \one{Each of the two subdomains is of size $165 \times 320 \times 320$ wavelengths}. \two{We also run the same simulation on 3 and 4 GPUs, split over as many subdomains (over the x dimension), to demonstrate the scalability of the domain decomposition framework.}

\begin{figure}
    \centering
    \includegraphics[width=\linewidth]{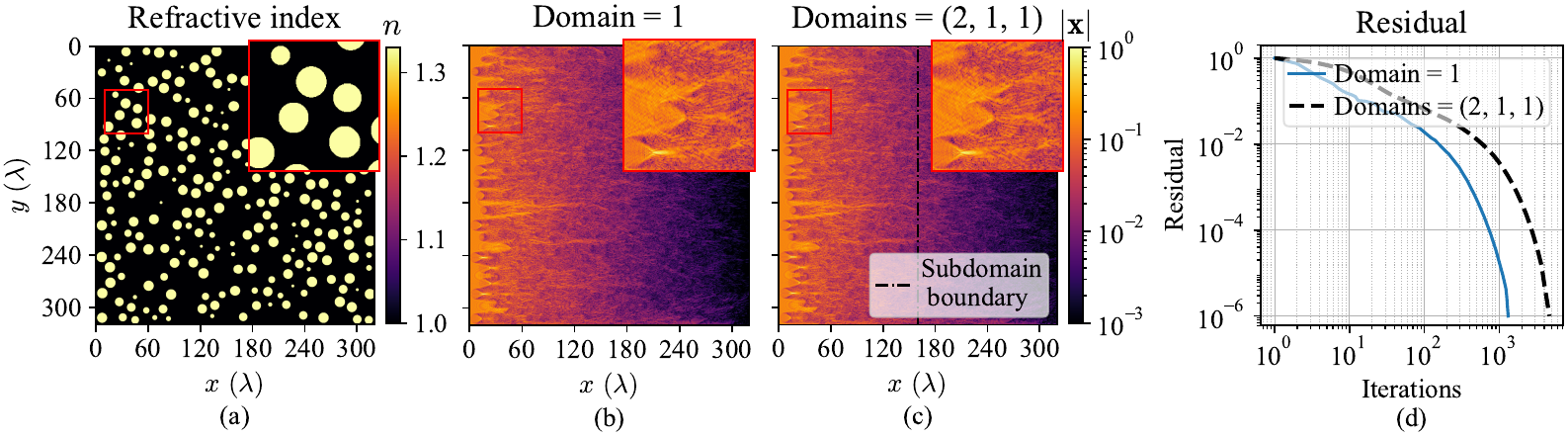}
    \caption{Large 3D simulation of size $320 \times 320 \times 320$ wavelengths with and without domain decomposition. A 2D slice of (a) the refractive index map, (b) the result from a 1-domain simulation on the CPU, and (c) the result from a 2-domain simulation using two GPUs, with the dash-dotted black line indicating the subdomain boundary. The insets are zoomed-in portions of the selected area to show finer details. \two{The spheres appear to be of different sizes because the 2D slice takes a cross-section of the spheres at different points. }(d) Residual as a function of the iterations for (b), a solid line, and (c), a dashed line.}
    \label{fig:dd_large_simulation}
\end{figure}

Fig.~\ref{fig:dd_large_simulation}a shows a slice of the 3D refractive index map through the centre of the z-axis without absorbing boundaries. The single-domain and 2-domain simulation results are shown in Figs.~\ref{fig:dd_large_simulation}b and c, respectively. The relative error between the two is $2.9\cdot 10^{-4}$. The insets in Fig.~\ref{fig:dd_large_simulation}a-c show finer details in a zoomed-in region of the slices.

Fig.~\ref{fig:dd_large_simulation}d shows the residual as a function of the iterations for the two simulations. The two-domain simulation (Fig.~\ref{fig:dd_large_simulation}c) that is run on 2 GPUs converges in 4697 iterations in $45$ minutes (2730 s), while the single-domain simulation (Fig.~\ref{fig:dd_large_simulation}b) converges in 1316 iterations in $15.5$ hours (56040 s) as it is run on the CPU (Table~\ref{tab:dd_large_simulation}). 

In the 2-domain case, the subdomain communication and removal of wrapping artefacts lower the scaling coefficient $c$, increasing the number of iterations by $3.5 \times$. \two{The 2-domain simulation is $20\times$ faster than the 1-domain simulation on the CPU. However, the overhead of the iteration increase is a trade-off to be considered.} In addition to this increase in the number of iterations, the domain decomposition framework has two overheads: the overhead of computing the edge corrections and the overhead due to data transfer between GPUs when the edge corrections are transferred and applied to neighbouring subdomains in each iteration. This subdomain communication leads to a lock-step execution, meaning the GPUs wait for the results of the other GPU(s). This overhead adds approximately $40\%$ time per cubic wavelength and reduces for larger simulations.

\begin{table}
    \centering
    \small
    \begin{tabular}{l|C{2.5cm}|C{1cm}|C{2cm}|c}
        \bfseries Run parameters & \bfseries Total size (cubic wavelengths) & \bfseries Time (s) & \bfseries Time $\cdot$ \#GPU (s) & \bfseries Iterations\\
        \hline\hline
        1 domain, CPU & $3.27\cdot 10^7$ & 56040 & & 1316 \\
        \hline
        2 domains, 2 GPUs & $3.27\cdot 10^7$ & 2730 & \two{5460} & 4697 \\
        \two{3 domains, 3 GPUs} & \two{$3.27\cdot 10^7$} & \two{2022} & \two{6066} & \two{4697} \\
        \two{4 domains, 4 GPUs} & \two{$3.27\cdot 10^7$} & \two{1600} & \two{6400} & \two{4697} \\
    \end{tabular}
    \caption{\two{The largest possible simulation with domain decomposition on 2 GPUs, with simulation runs on the CPU for comparison, and 3 and 4 GPUs to illustrate scalability. The simulations were run on a system with two Silver-4216 2.10 GHz CPUs with 128 GB RAM and four A40 48GB GPUs. The domain decomposition simulations on the GPUs (rows 2-4) take $3.5 \times$ the number of iterations for convergence as the CPU simulation (row 1), but are at least $20\times$ faster. Domain decomposition reduces the total computation time (column 3), indicating parallel execution, and adds only about 10\% overhead to the total amount of GPU-seconds (column 4) for each GPU added, while leaving the total number of iterations constant (column 5) from 2 domains onwards.}}
    \label{tab:dd_large_simulation}
\end{table}

Despite this overhead, the 2-domain, 2-GPU simulation is $20 \times$ faster than the single-domain simulation on the CPU without domain decomposition. With these two GPUs, we increased the simulation size by $1.95 \times$ of what can be accommodated on a single GPU to an unprecedented $320 \times 320 \times 320$ wavelengths. \one{When the domain is split into more subdomains along a single axis, there is no additional overhead in terms of iterations beyond the initial step of going from one to two subdomains, as seen from the last column in Table~\ref{tab:dd_large_simulation}. When more GPUs are used, the total simulation time reduces, indicating a parallelisation of the computations. The total number of GPU-seconds used for the simulation only increases by about 10\% overhead per GPU, indicating a small synchronisation and communication overhead.}

\section{Conclusion}
\label{sec:conclusions}
\noindent We have introduced a domain decomposition of the modified Born series (MBS) approach \cite{osnabrugge2016convergent, vettenburg2023universal} applied to the Helmholtz equation. With the new framework, we simulated a complex 3D structure of a remarkable $3.27\cdot 10^7$ wavelengths in size in just $45$ minutes by solving over two GPUs. This result is a factor of $1.95$ increase over the largest possible simulation on a single GPU without domain decomposition. \two{Furthermore, we showed that excellent scaling is maintained up to at least four GPUs.} 

Our decomposition framework hinges on the ability to split the linear system as $A=L+V$. Instead of the traditional splitting, where $V$ is a scattering potential that acts locally on each voxel, we introduced a $V$ that includes the communication between subdomains and the removal of wraparound artefacts. As a result, the operator $(L+I)^{-1}$ in the MBS iteration can be evaluated locally on each subdomain using a fast convolution. Therefore, this operator, the most computationally intensive step of the iteration, can be evaluated in parallel on multiple GPUs. 

With the current 2-GPU simulation, we were able to solve a problem of $320 \times 320 \times 320$ wavelengths $20 \times$ faster than without domain decomposition, as the non-decomposed problem is too large to fit on a single GPU. \two{This marks a significant enhancement to the MBS in terms of performance and the scalability of simulations. However, there are two trade-offs associated with this gain in speed and simulation size due to the overheads associated with our domain splitting method. First, there is a significant overhead in terms of an increased number of iterations to solve the problem with multiple domains than with a single domain. Second, there is also the communication and synchronisation overhead, but} as we demonstrated in Section~\ref{subsubsec:results_subdomains}, there is only a slight overhead associated with adding more subdomains along an axis after the first splitting. This favourable scaling paves the way for distributing simulations over more GPUs or compute nodes in a cluster.

In this work, we have already introduced strategies to reduce the overhead of the domain decomposition through truncating the blocks for subdomain communication and wrapping artefacts correction to only a few points close to the edge of the subdomain, and only activating certain subdomains in the iteration. We anticipate that further developments and optimisation of the code may help reduce the overhead of the lock-step execution. 

Finally, due to the generality of our approach, we expect it to be readily extended to include Maxwell's equations \cite{kruger2017solution} and birefringent media \cite{vettenburg2019calculating}. Given the rapid developments of GPU hardware and compute clusters, we anticipate that optical simulations at a cubic-millimetre scale can soon be performed in a matter of minutes.

\section{Code availability}
\noindent The open-source Python implementation of our method is available on GitHub \cite{code_wavesim_python}.

\section{Acknowledgements}
\noindent This work was supported by the ERC Proof of Concept grant WAVESIM (number 101069402) and a Vici grant (number 21646) by the Dutch Research Council (NWO). We thank Vassilis Sarantos, Evangelos Marakis, Michael Ktistakis, Ioannis Zaharakis, Vasilis Papadopoulos and Tom Vettenburg for motivating discussions, and Arnon A.B. for proofreading the manuscript.

%
%%%%%%%%%%%%%%%%%%%%%%% References %%%%%%%%%%%%%%%%%%%%%%%
\bibliographystyle{elsarticle-num} 
\bibliography{wavesim}

\end{document}